\documentclass[referee]{raa}            

\usepackage{graphicx,times}             
\usepackage{natbib}
\usepackage{amssymb,amsmath}
\usepackage{booktabs,subcaption}
\usepackage{multirow,threeparttable}
\usepackage{color}
\bibpunct{(}{)}{;}{a}{}{,}

\usepackage[a4paper=true,pagebackref=true]{hyperref}
\hypersetup{colorlinks = true, linkcolor = green, anchorcolor = red, citecolor = blue, filecolor = red, pagecolor = red, urlcolor = red}

\begin{document}

   \title{ Predicting the Evolution of Photospheric Magnetic Field in Solar Active Regions Using Deep Learning
}

   \volnopage{Vol.0 (20xx) No.0, 000--000}      
   \setcounter{page}{1}          
   \author{Liang Bai \inst{1} \and Yi Bi \inst{2} \and Bo Yang \inst{2} \and Jun-Chao Hong \inst{2}
   \and Zhe Xu \inst{3} \and Zhen-Hong Shang \inst{1,4} \and Hui Liu \inst{2} \and Hai-Sheng Ji \inst{3} \and Kai-Fan Ji\footnotetext{ $\star$ Corresponding author} \inst{2,\star} }

   \institute{
        Faculty of Information Engineering and Automation, Kunming University of Science and Technology, Kunming  650500  \\
        \and Yunnan Observatories, Chinese Academy of Sciences, Kunming 650216; {\it jkf@ynao.ac.cn}\\
        \and Purple Mountain Observatory, Chinese Academy of Sciences, Nanjing  210034\\
        \and Yunnan Key Laboratory of Artifical Intelligence, Kunming University of Science and Technology, Kunming  650500\\
\vs\no
   {\small Received~~20xx month day; accepted~~20xx~~month day}}

\abstract{
   The continuous observation of the magnetic field by Solar Dynamics Observatory (SDO)/Helioseismic and Magnetic Imager (HMI) produces numerous image sequences in time and space. These sequences provide data support for predicting the evolution of photospheric magnetic field. Based on the spatiotemporal long short-term memory(LSTM) network, we use the preprocessed data of photospheric magnetic field in active regions to build a prediction model for magnetic field evolution. Because of the elaborate learning and memory mechanism, the trained model can characterize the inherent relationships contained in spatiotemporal features. The testing results of the prediction model indicate that (1) the prediction pattern learned by the model can be applied to predict the evolution of new magnetic field in the next 6 hour that have not been trained, and predicted results are roughly consistent with real observed magnetic field evolution in terms of large-scale structure and movement speed; (2) the performance of the model is related to the prediction time; the shorter the prediction time, the higher the accuracy of the predicted results; (3) the performance of the model is stable not only for active regions in the north and south but also for data in positive and negative regions. Detailed experimental results and discussions on magnetic flux emergence and magnetic neutral lines finally show that the proposed model could effectively predict the large-scale and short-term evolution of the photospheric magnetic field in active regions. Moreover, our study may provide a reference for the spatiotemporal prediction of other solar activities. 
}
   \authorrunning{L. Bai et al. }   
   \titlerunning{Predict Evolution of Photospheric Magnetic Field}  

   \maketitle
%
%
\section{Introduction}           
\label{sect:intro}


The ability to predict future results is a key component of intelligent decision-making systems(\citealt{oprea_review_2020_1}). In recent years, with the in-depth research and widespread application of artificial intelligence and machine learning, various prediction and forecasting algorithms have been proposed. For example, many studies, such as  \cite{shi_convolutional_nodate_2}, \cite{kwon_predicting_2019_3}, \cite{wang_memory_2019_4}, have provided new developments and phased breakthroughs for short-term video prediction. Due to the complicated spatiotemporal relationships and uncertainty between video frames, video prediction is a very challenging task. However, machine learning algorithms represented by deep learning can dig out deterministic relationships from a large amount of data including uncertainty. At the same time, the spatial semantics in videos also have a strong correlation and continuity in the transformation and movement of the time dimension. Through the self-supervised statistical learning on large amounts of data, the specially designed neural network can better mine the deterministic temporal and spatial relationships in videos, and then realize the short-term prediction of the future frames.

The application of machine learning algorithms in solar activity research mostly focuses on the prediction of some solar activity parameters. For example, \cite{nishizuka_deep_2018_6} used a neural network model to predict the probability of solar flare occurrence and the maximum flow level. \cite{Huang_2018} used a deep learning model to automatically mine solar flare forecasting patterns from solar line-of-sight magnetograms. \cite{pala_forecasting_2019_5} used a recurrent neural network model to predict the number change  of sunspots over time. \cite{dani_prediction_2019_7} used a machine learning model to predict the maximum amplitude of solar cycle 25. Recently, there were also studies on predicting the distribution of sunspot butterfly diagrams(\citealt{covas_neural_2019_8}) and the solar surface longitudinally averaged unsigned radial component magnetic field(\citealt{covas_transfer_2020_9}). These studies have greatly improved our understanding for the rules of various solar activities. However, there is no work to predict the evolution of the photospheric magnetic field in active regions(AR). Photospheric magnetic field is usually generated in the solar interior,  and controls most physical processes in solar atmosphere(\citealt{wiegelmann_magnetic_2014_10}). The photospheric magnetic field in active regions often has higher field strength(\citealt{getachew_spatial-temporal_nodate_11}), which is not only closely related to various magnetic activities that occur in the solar atmosphere, such as sunspots, solar wind, and coronal mass ejection, etc., but also affects our Earth's near-space environment, climate changes, and daily life. Therefore, research on predicting the evolution of the photospheric magnetic field in active regions is of great significance.

Inspired by the success of video prediction researches in the field of computer vision, we guess that the magnetic field evolution in active regions may also contain some deterministic relations that can be predicted. Meanwhile, along with the entire life cycle of sunspots, the evolution of photospheric magnetic fields provides many continuous image sequences in time and space, just like frames of video. Hence, we try to build the prediction model of magnetic field evolution using a neural network. We construct a magnetic field evolution data set as complete as possible and train our model with a spatiotemporal LSTM network. Specifically, we make some modifications to the memory in memory(MIM) network(\citealt{wang_memory_2019_4}) to make it suitable for the spatiotemporal prediction of the magnetic field evolution in active regions. After training, we test the effect of the trained model on the whole test set and give detailed analyses. The experimental results verify our conjecture: the trained model could indeed mine the complex spatiotemporal physical relationships contained in the training set of magnetic field evolution. Inputting a new 12-hour magnetic field sequence, this model can output the predicted large-scale magnetic field evolution in the next 6 hours. This prediction ability fully shows that our model has learned a prediction pattern from a large number of training set, which can be applied to predict the evolution of new magnetic field that have not been trained. When the new 12-hour magnetic field images are input, their spatial features are extracted by convolution operations, and the evolutionary relationships between them are learned by the memory network with recurrent structure. Finally, combined with the extracted spatiotemporal features, the learned prediction pattern outputs the predicted results in a recurrent way. It should be pointed out that the model is mainly used for large-scale prediction of solar magnetic field evolution, and the prediction is entirely kinematic prediction without considering dynamic factors. When it comes to the studies of the quiet Sun region, the fine structure evolution of magnetic field, and dynamics, the results acquired may be not reliable. We believe this is also of great significance to the macroscopic study of the photospheric magnetic field evolution in active regions.

The article is organized as follows: The data is described in Section 2. The network and experiment settings are shown in Section 3. The results and analysis of predicted results for magnetic field evolution are given in Section 4. Finally, conclusions and discussions are drawn in Section 5.
\section{Data}
\label{sect:data}
The HMI on board the SDO provides the accurate measurement of solar photospheric magnetic field so far(\citealt{pesnell_solar_2012_12}). The Space-weather HMI Active Region Patches (SHARPs) data by HMI records images in patches that encompass automatically tracked magnetic concentrations for their entire lifetime with a 12-minute cadence(\citealt{bobra_helioseismic_2014_13}). In this paper, we mainly use the radial component of the vector magnetic field in a heliographic Cylindrical Equal-Area (CEA) coordinate system. We collate the magnetic field evolution data of the solar active regions from 2011 to 2015. The magnetic field images with longitudes outside ±60° of the central meridian are excluded to avoid the influence of projection effects. Active regions whose evolution time is less than 24 hours are also discarded so that the obtained data can form sequences. The numbers of active regions in the south and the north are kept as close as possible. After these data selection criterias, 46 active regions were obtained, of which 26 were located in the northern hemisphere and 20 in the southern hemisphere.

When using 12 minutes as the interval of the sequence, the change between every two frames of the magnetic field images is small, and the predicted results often fail to better reflect the superiority of the model. So, we set the time interval of magnetic field sequence to 1 hour, that is, one frame is sampled every 5 frames. In order to improve the accuracy of predicted results, we also perform the following preprocessing operations on magnetic field data. Firstly, the discontinuous parts of the magnetic field sequences are removed. Secondly, to suppress the interference of noise, the part where absolute value of magnetic field strength is lower than 300 Gauss is removed, and the range of absolute field strength is kept in [300,3000] Gauss. Thirdly, to increase the learning efficiency of the model, the data range is normalized to [-1,1], and the magnetic field images are down-sampled using the binning method(\citealt{li_image_2009_14}). Fourthly, training set is enhanced by using the flip of positive and negative regions, the flip of left-right, to enhance the generalization of the model during training process. At the end, all magnetic field images are cut into sequence of 18 consecutive frames.

Finally, the preprocessed data is randomly divided into a training set and a testing set. The obtained training set covers 35 active regions with 3603 magnetic field evolution sequences, and the testing set includes 11 active regions with 621 sequences in total. The predicted results outputted by the model need to be de-normalized to ensure that the data range is consistent with the real magnetic field sequences.


\section{Model}
\label{sect:Obs}
\subsection{Network}
Supposing $X_{t}\in R^{w\times h}$  is the $t$-th frame in the magnetic field sequence $X=(X_{t-n},\cdots ,X_{t-1},X_{t})$, where $w$ and $h$ denote width and height of the image respectively. Our goal is to predict the forthcoming magnetic field sequence $Y=(\hat{Y}_{t+1},\hat{Y}_{t+2},\cdots ,\hat{Y}_{t+m} )$. As shown in Figure 1, the spatiotemporal LSTM neural network used is composed of one spatiotemporal LSTM module (ST-LSTM) and two MIM module. This network comes from MIM network(\citealt{wang_memory_2019_4}), and we just make some modifications. Input a continuous 12-hour magnetic field sequence, this network outputs a predicted magnetic field evolution in the next 6 hours. In the training process, we use the mean square error between the 6-hour prediction results and the real magnetic field images as the loss function. Owing to the recurrent network structure, the trained model can fully integrate the temporal and spatial features. As a result, long-term and short-term spatiotemporal memorys are established, which can better guide the prediction of the magnetic field evolution.

\begin{figure} 
  \centering
  \includegraphics[width=12.0cm, angle=0]{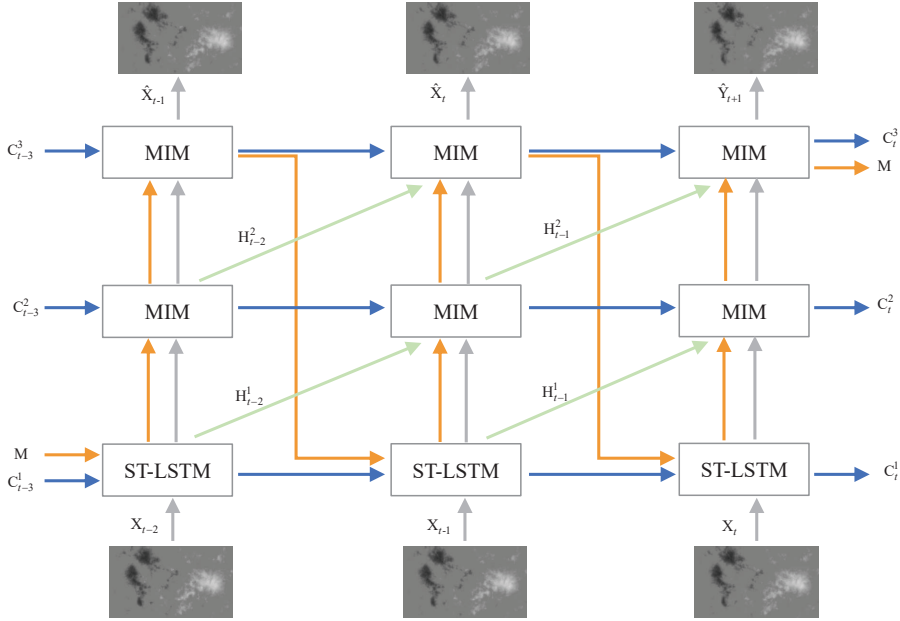}
  \caption{The network structure we used is shown. The ST-LSTM module and MIM module have their own memory state (C in Figure) and a shared spatial memory state (M in Figure). These memory states could selectively memorize and store the input magnetic field information or high-dimensional abstract middle features. The recurrent network structure realizes the horizontal flow of the memory state of each module (blue arrows in Figure) and the zigzag flow of the shared spatial memory state M (yellow arrows in Figure) between modules. } 
  \label{Fig1}
\end{figure}

The evolution process of the photospheric magnetic field in active regions includes the movement, deformation, and emergence of the magnetic field flux patches. A sequence may contain both deterministic changes such as the movement of magnetic field structures and non-deterministic changes such as the deformation and emergence. This brings great challenges for predicting the magnetic field evolution. But, the network we used can deal with these problems to a certain extent. In Figure 1, the flow of features in different directions promotes the better integration of the temporal and spatial features of the magnetic field sequence. Inputing the hidden state of the previous module at the previous moment (light green arrows in Figure 1) and the hidden state of the current moment at the same time, the MIM module could roughly characterize the deterministic evolution relationships. In the process of implementation, we change the hadamard product of the state update in the standard ST-LSTM and MIM modules to convolution operations. This not only reduces the parameter scale of the network under the same settings, but also makes the network more flexible without limiting the size of the input magnetic field images. In addition,  it is believed that the gate structure of different functions in each module should input independent distributed data, rather than input the same distributed data. So we replace the layer normalization(\citealt{ba_layer_2016_15}) in all modules with group normalization(\citealt{wu_group_2018_16}).

\subsection{ Experimental Settings}

Referencing to the official code based on TensorFlow\footnote{https://github.com/Yunbo426/MIM} and incomplete open source code\footnote{https://github.com/coolsunxu/MIM\_Pytorch}, we reimplemented a new version of the network  derived from the PyTorch framework(\citealt{paszke_pytorch_nodate_17}). Our code is available online\footnote{https://www.github.com/beiyan1911/magnetic-field-predict}. Our model is trained and tested on the Windows 7 operating system. The hardware environment consistents of Intel Xeon E5-2650 CPU, NVIDIA Tesla P100 GPU with 16G memory, 16G RAM, etc. We use Compute Unified Device Architecture(CUDA) 10.1 and CUDA Deep Neural Network library(cuDNN) 7.0 to implement GPU acceleration for this model, and use Adam optimizer(\citealt{kingma_adam_2017_18}) to update the  parameters. The hyperparameters are set to $\beta_{1} = 0.9$, $\beta_{2} = 0.999$, learning rate is 0.001, batch size is 1. As well as \cite{wang_memory_2019_4}, we use the guiding learning strategy(\citealt{bengio_scheduled_nodate_19}) to train the model to improve its robustness.

\section{results and analysis}
\label{sect:analysis}

\subsection{Analyses on Large Scale Structure}
\begin{figure}[htbp]
   \label{fig:example}
   \centering
   \begin{subfigure}{\textwidth}
      \centering
      \includegraphics[width=\textwidth]{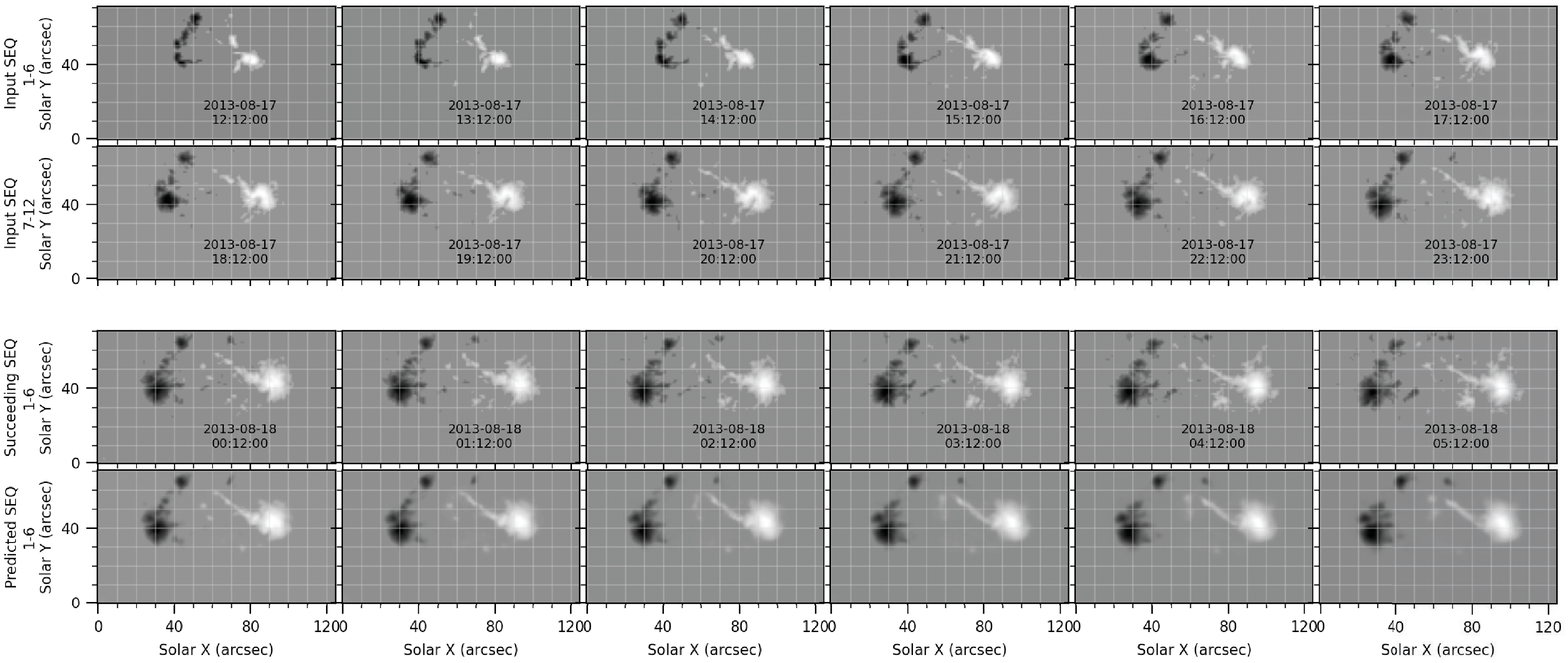}
      \caption{One Group of Predicted Results on AR 11824}
   \end{subfigure}

   \begin{subfigure}{\textwidth}
      \centering
      \includegraphics[width=\textwidth]{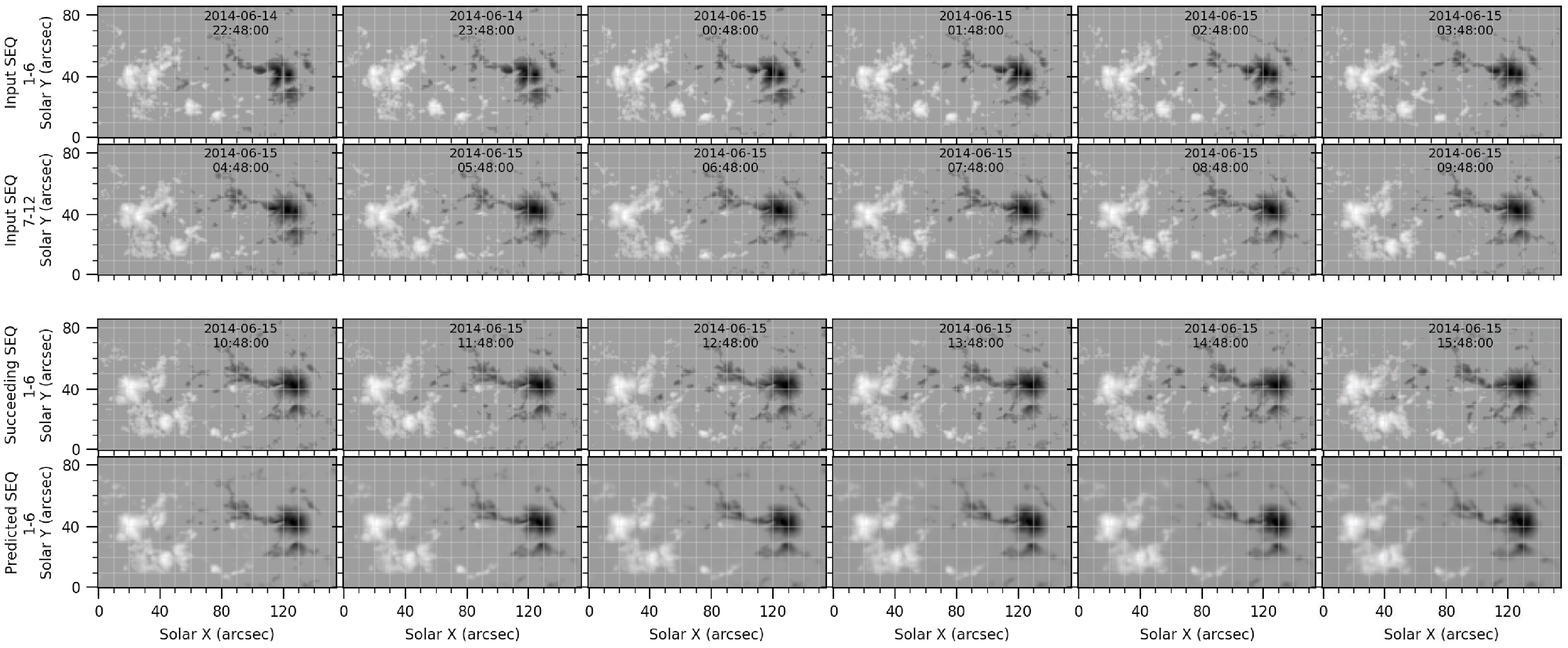}
      \caption{One Group of Predicted Results on AR 12089}
   \end{subfigure}
   \caption{Two groups of predicted results are shown. In panels(a-b), rows 1 and 2 are the input 12-hour magnetic field sequence, the third row is the real magnetic field sequence 6 hours after the input sequence, and the fourth row is the predicted magnetic field evolution by our model for the next 6 hours. For comparison, the magnetic field images are gridded.}
\end{figure}

In order to show the consistency between predicted results and the real observed magnetic field images, we select 2 groups of predicted results from AR 11824 and AR 12089 to display. In panels(a-b) of Figure 2, by comparing the 3rd and 4th rows of the magnetic field images, it can be seen that the predicted results are generally consistent with the real magnetic field images in the large-scale magnetic field structure. Moreover, with the increase of the prediction time (from left to right in Figure 2), the predicted results are generally coincident with the real magnetic field sequence in large-scale changes. Although, the predicted results are different from the real magnetic field images in the shape of some fine magnetic structure.

We calculate three metrics: Correlation Coefficient (CC), Structural Similarity (SSIM), and Root Mean Square Error (RMSE), which can quantitatively evaluate the degree of consistency between the predicted results and the real magnetic field images. CC can be uesd to evaluate the linear correlation between the two magnetic field images in field strength. Its value is in the range of [- 1,1]. The closer the value to - 1, the stronger the negative correlation between the two images. The closer the value to 1, the stronger the positive correlation. SSIM is used to calculate the structural similarity of two images. Its value is in the range of [0,1]. The larger the value, the higher the similarity. RMSE can evaluate the numerical difference between two images. The greater its value, the greater the numerical difference. The formula of SSIM is as follows:
\begin{equation}
   SSIM(x,y)=\frac{(2u_{x}u_{y}+c_{1} )(2\sigma_{xy}+c_{2})}{(u_{x}^{2}+u_{y}^{2}+c_{1})(\sigma_{x}^{2}+\sigma_{y}^{2}+c_{2}) } 
\label{eq:LebsequeI}
\end{equation}
where $x$ and $y$ represent two magnetic field images, $u_{x}$ and $u_{y}$ are the arerages of $x$ and  $y$, $\sigma_{x}^{2}$ is the variance of $x$, $\sigma_{y}^{2}$ is the variance of $y$, $\sigma_{xy}$ is the covariance of $x$ and $y$, $c_{1}$ and $c_{2}$ are two variables to stabilize the division with weak denominator.

\begin{table}
   \begin{center}
   \caption[]{ 
      Three Metric Values for Two Groups of Predicted Results  \quad \quad }
   \label{Tab:publ-works}
   \begin{threeparttable}      
   \begin{tabular}{cccccccc}
   \toprule
   Examples &	Metrics&	T = 1	& T = 2 & T = 3 & T = 4 & T = 5 & T = 6                    \\
   \midrule
   \multirow{3}{*}{AR 11824}   & SSIM	&0.94   &0.90   &0.86   &0.82   &0.79   &0.77  \\
                              & CC	   &0.97   &0.95   &0.92   &0.90   &0.87   &0.84   \\
                              & RMSE	&80.56  &109.28 &137.22 &159.92 &179.87 &197.95 \\
   \midrule
   \multirow{3}{*}{AR 12089}  & SSIM	&0.86   &0.78   &0.72   &0.68   &0.65   &0.63   \\
                             &CC	   &0.97   &0.94   &0.91   &0.89   &0.88   &0.87   \\
                             & RMSE	&117.20 &158.79 &187.94 &211.12 &220.96 &227.95 \\  
   \bottomrule
   \end{tabular}
   \begin{tablenotes}   
      \footnotesize         
      \item[] T = 1,2,...,6 represents the hour length of the predicted result, and the unit of RMSE is Gauss.
    \end{tablenotes} 
   \end{threeparttable}
   \end{center}
\end{table}

\begin{figure}[htbp]
   \label{fig:example}
   \begin{subfigure}{\textwidth}
      \centering
      \includegraphics[width=\textwidth]{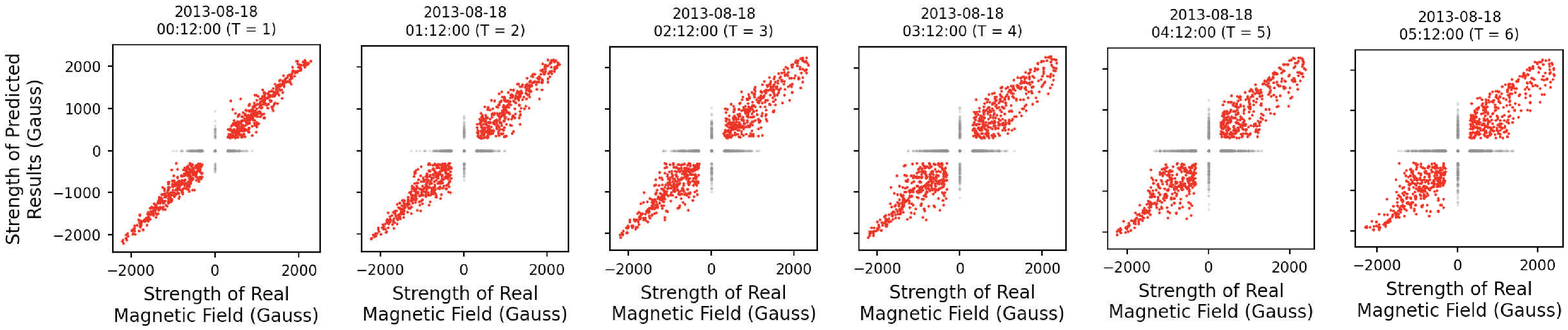}
      \caption{One Group of Correlation Diagrams on AR 11824}
   \end{subfigure}

   \begin{subfigure}{\textwidth}
      \centering
      \includegraphics[width=\textwidth]{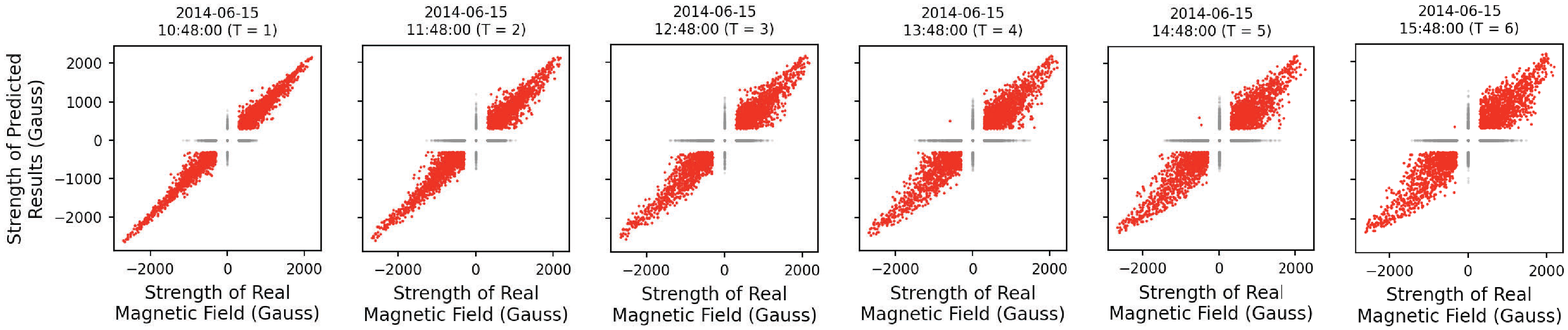}
      \caption{One Group of Correlation Diagrams on AR 12089}
   \end{subfigure}
   \caption{The correlation diagrams of the two examples are shown. Except for the background field strength of 0, the predicted and real magnetic field data do not contain the area with the absolute field strength less than 300 Gauss. So, there are two parts in the correlation diagrams. The red scatter points indicate the intersection area where the absolute field strength of the predicted and the real magnetic field images is greater than 300 Gauss. The light gray scatter points indicate other areas, that is, where the structure of the predicted results is different from that of the real magnetic field.}
\end{figure}

Table 1 shows the three metrics calculated on the two groups of prediction results (corresponding to the two examples in Figure 2). Figure 3 shows correlation diagrams of the two groups of examples between the predicted and real magnetic field images. We can clearly see from Table 1 and Figure 3 that with the increase of prediction time, the CC and SSIM metrics of the two examples show a downward trend. The numerical error RMSE continues to increase, and the dispersion of the correlation diagrams becomes larger and larger. These changes indicate that the accuracy of predicted results gradually decreases with the increase of prediction time.

To better evaluate the average degree of consistency between the predicted and the real magnetic field images, we also calculate the three metrics on the whole testing set. Figure 4 shows the average, and the range of standard deviation, of the three metrics. From the changes in panels(a-c) of Figure 4, it can be seen that as the prediction time increases, the average CC between the predicted results and the real magnetic field images gradually decreases from 0.96 to 0.83. The average SSIM gradually decreases from 0.92 to 0.76. The average error RMSE gradually increased from 85.64 Gauss to 179.17 Gauss. The standard deviation(height of error bar in Figure 4) of the three metrics also gradually increased. These phenomena show that the average accuracy of prediction results on the whole testing set also gradually decreases with the increase of prediction time. But even in the last frame of the predicted results, the averages of CC and SSIM values are  0.83 and 0.76, which shows that the last frame of the predicted results are still roughly consistent with the real magnetic field images. Therefore, we suggested that our model could predict the large-scale evolution of magnetic field sequence in the next 6 hours.

\begin{figure}[htbp]
   \label{fig:example}
   \begin{subfigure}{0.3\textwidth}
      \centering
      \includegraphics[width=\textwidth]{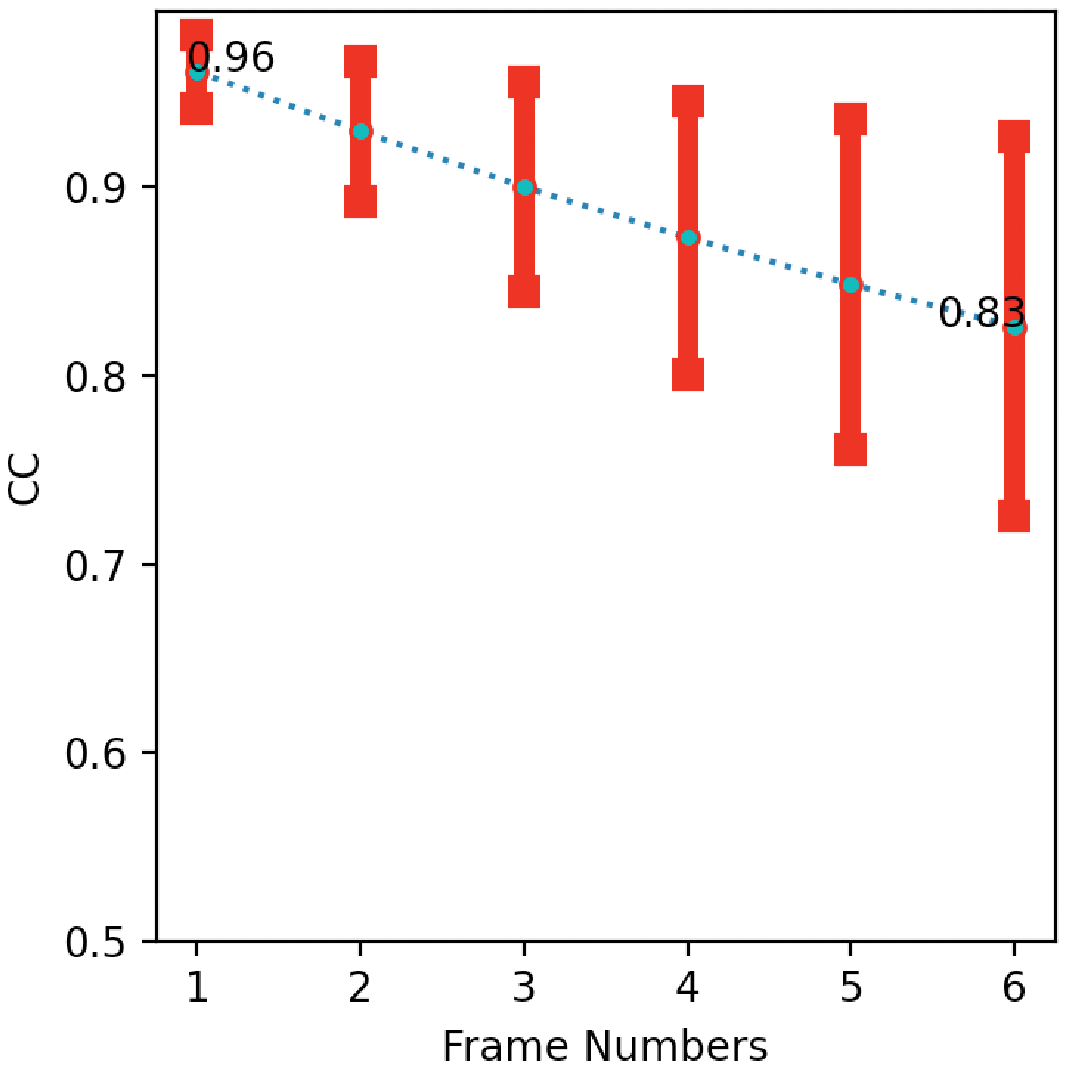}
      \caption{Average of CC}
   \end{subfigure}
   \begin{subfigure}{0.3\textwidth}
      \centering
      \includegraphics[width=\textwidth]{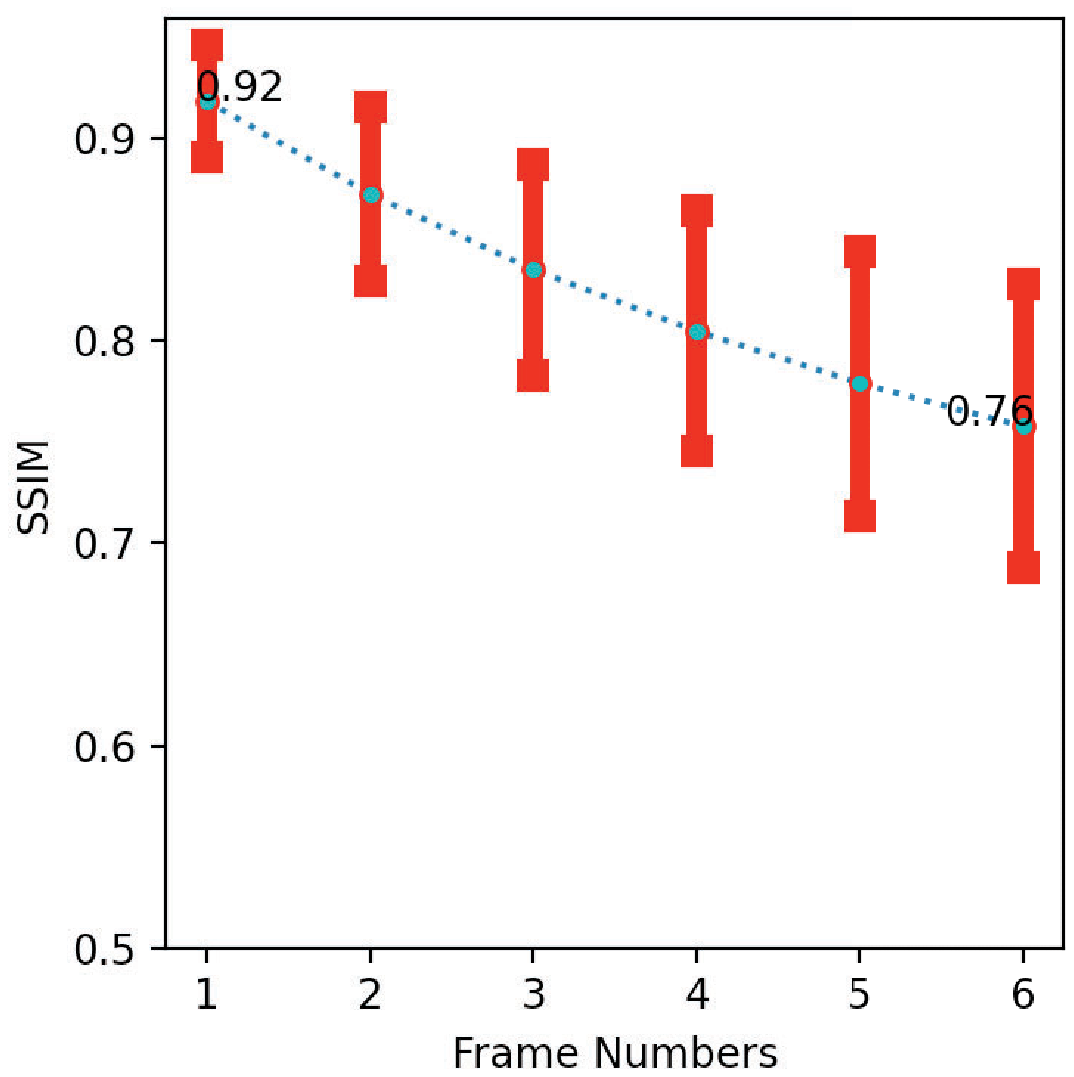}
      \caption{Average of SSIM}
   \end{subfigure}
   \begin{subfigure}{0.3\textwidth}
      \centering
      \includegraphics[width=\textwidth]{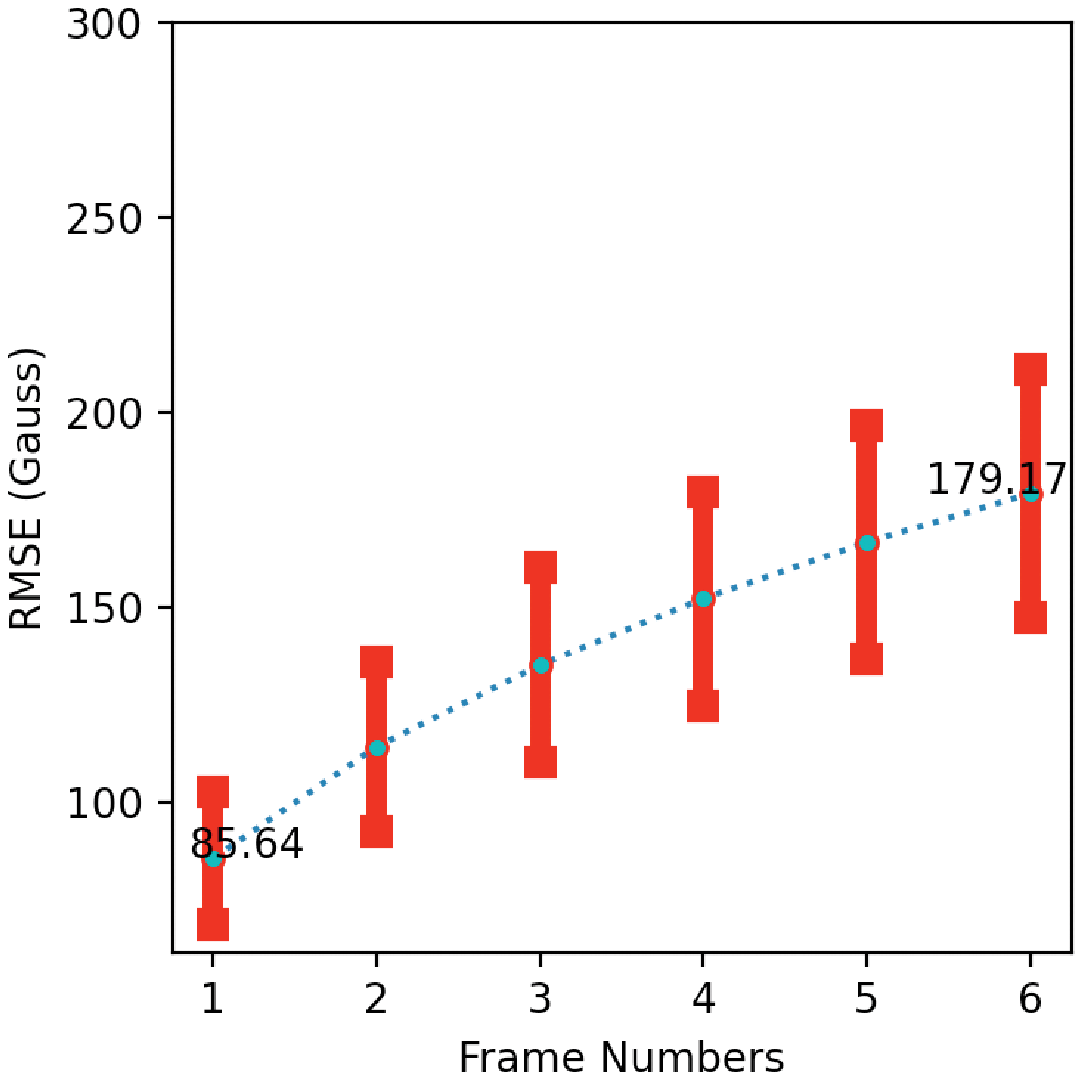}
      \caption{Average of RMSE}
   \end{subfigure}
   \caption{The Average of CC, SSIM, and RMSE on the Whole Testing Set}
\end{figure}

In Figures 2 and 4, the example of AR 11824 is in the southern hemisphere, and the example of AR 12089 is in the northern hemisphere. We compare not only these two groups of examples but also more predicted results during the experiment, and find that the predicted results are relatively stable in different positions and different positive and negative regions. Specifically, there is no difference in predicted results between the positive regions and the negative regions. And there is also no difference in predicted results between active regions in the north and south. These are in line with the characteristics of the model based on deep learning: the stability of the model performance is related to whether the training data set is complete or not. The more complete the training data, the better the performance of the model on the same type of new data. In the data preparation stage, we try to make the training set cover a similar number of active regions in the north and south. Data enhancement methods, such as positive and negative flip, left-right flip, are also used to avoid the influence of uneven data distribution on the prediction results.

The accuracy of the predicted results decreases with the increase of prediction time, which is in line with our expectation. Because the network structure of the model is a recurrent neural network, the output of each predicted image needs to input the previous frame and the hidden state. When performing prediction, the input image is the predicted result of the previous frame, and the error of the previous predicted result will gradually accumulate into the subsequent predicted frame. So, with the increase in the number of prediction frames, the cumulative error of predicted results gradually increases.

\subsection{Analyses on Movement Speed}
Another important aspect in the evolution of the magnetic field is the movement of the magnetic structure. We also analyse the large-scale movement speed of the photospheric magnetic field in active regions. Specifically, we draw the movement speed diagrams about the two groups of predicted results (corresponding to the two examples in Figure 2) by using the optical flow method(\citealt{goos_two-frame_2003_20}). As shown in Figure 5, to display the speed of the large-scale magnetic field structure more clearly, we only show the union area where the absolute field strength is greater than 300 Gauss between the real and predicted magnetic field images. Besides, we also show the results more intuitively in the form of animation, which can be seen online\footnote{https://github.com/beiyan1911/magnetic-field-predict/tree/main/animation}. With the increase of the prediction time in Figure 5, we can see that the directions and magnitudes of predicted velocity in the longitude direction are generally consistent with the real speed. In the latitude direction, the directions and magnitudes of the predicted speed and the real speed are also roughly consistent. In some small areas, the predicted speed directions are opposite to the real speed directions, and the speed magnitudes are also different. Meanwhile, in both longitude and latitude directions, with the increase of the prediction time, changes of the speed magnitudes and directions on the predicted and real speed are very small, which also verify that the movement of the magnetic structure in evolution processes has good coherence and consistency. Through above analyses, it can be concluded that the predicted results are generally consistent with the real magnetic field sequence in the large-scale movement speed.

\begin{figure}[htbp]
   \label{fig:example}
   \begin{subfigure}{\textwidth}
      \centering
      \includegraphics[width=\textwidth]{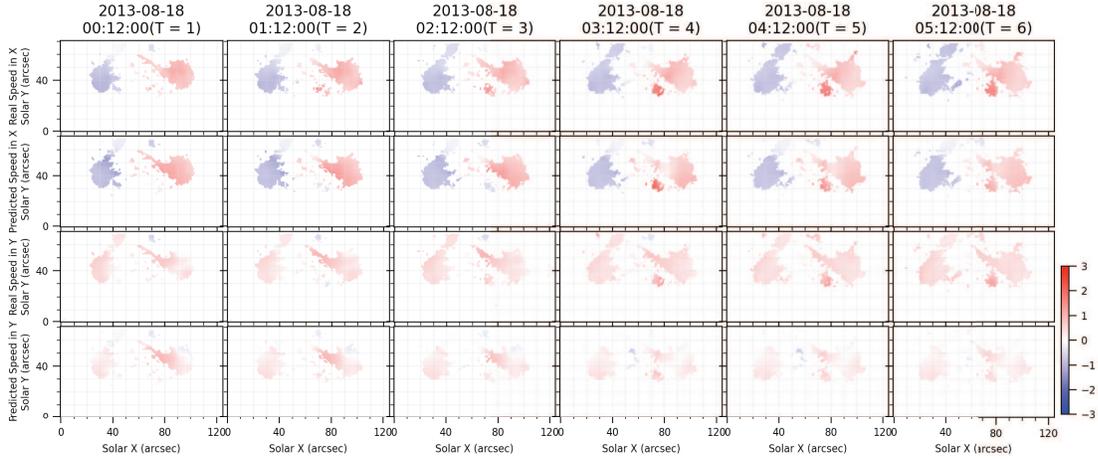}
      \caption{One Group of Speed Diagrams on AR 11824}
   \end{subfigure}

   \begin{subfigure}{\textwidth}
      \centering
      \includegraphics[width=\textwidth]{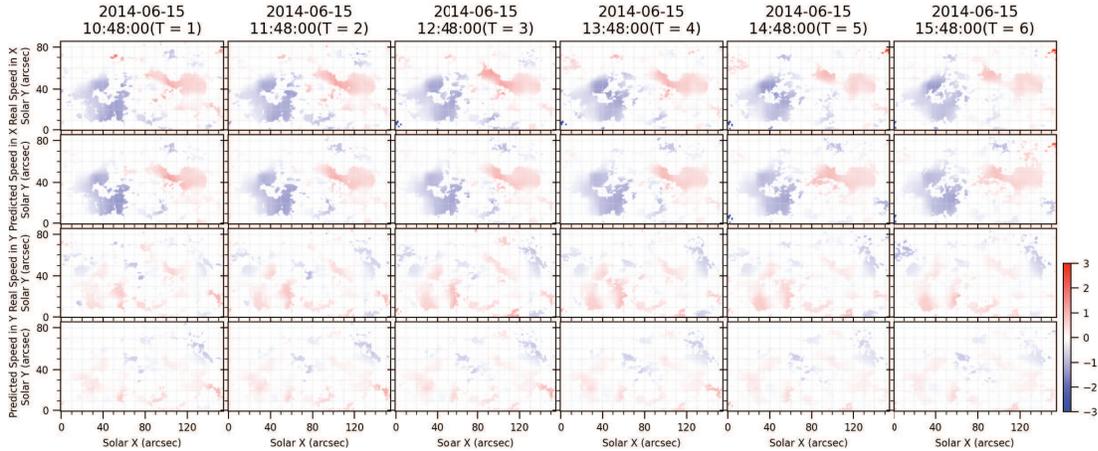}
      \caption{One Group of Speed Diagrams on AR 12089}
   \end{subfigure}
   \caption{The movement speed diagrams corresponding to the two groups of predicted results are displayed. In panels (a-b), the first and second rows are the real and predicted movement speed diagrams in the longitude direction (X direction) compared with the 12th input magnetic field image respectively. The 3rd and 4th rows are the real and predicted movement speed diagrams in the latitude direction ( Y direction) respectively. Red and blue patches in the speed diagrams indicate the opposite direction of movement, and the darker the color, the greater the speed magnitude. The unit of colorbar is arcsec/h.}
\end{figure}

 The SSIM and CC metrics of the real and predict speed diagrams on the two examples are shown in  Table 2. The average SSIM and CC change curves of the movement speed calculated on the entire test set are shown in Figure 6. It can be seen from Table 2 and Figure 6  that the SSIM and CC of the movement speed diagrams in the longitude direction are significantly higher than that in the latitude direction, regardless of whether it is on a single example or the statistical results of the whole test set. From Figure 6 we can also see that as the prediction time increases, the CC and SSIM metrics in the longitude and latitude directions generally show a downward trend, which is the same as the conclusion that the accuracy of the predicted results decreases as the prediction time increases in Section 4.1.
\begin{table}
   \begin{center}
   \caption[]{  SSIM and CC of Speed Diagrams in Different Active Regions}
   \label{Tab:publ-works}
   \begin{threeparttable}
   \begin{tabular}{cccccccc}
   \toprule
   Examples &	Metrics&	T = 1	& T = 2 & T = 3 & T = 4 & T = 5 & T = 6                    \\
   \midrule
   \multirow{4}{*}{AR 11824 }  & CC In X 	&0.95  &0.92 &0.86 &0.92 &0.90 &0.91\\
                              & CC In Y   &0.67  &0.66 &0.52 &0.47 &0.41 &0.40  \\
                              & SSIM In X	&0.58  &0.61 &0.53 &0.50 &0.52 &0.55 \\
                              & SSIM In Y	&0.39  &0.40 &0.36 &0.27 &0.26 &0.27 \\
   \midrule
   \multirow{4}{*}{AR 12089 }&CC In X 	&0.88 &0.90 &0.74 &0.78 &0.72 &0.82 \\
                            &CC In Y   &0.66 &0.71 &0.70 &0.64 &0.65 &0.53 \\
                            &SSIM In X	&0.64 &0.62 &0.68 &0.61 &0.59 &0.60 \\  
                            &SSIM In Y	&0.37 &0.35 &0.34 &0.31 &0.30 &0.27 \\  
   \bottomrule
   \end{tabular}
   \begin{tablenotes}   
      \footnotesize         
      \item[] These two examples still correspond to the two groups of predicted results in Figure 2. X represents the longitude direction, and Y represents the latitude direction.
    \end{tablenotes} 
   \end{threeparttable}
   \end{center}
\end{table}

\begin{figure}[htbp]
   \label{fig:example}
   \centering
   \begin{subfigure}{0.35\textwidth}
      \centering
      \includegraphics[width=\textwidth]{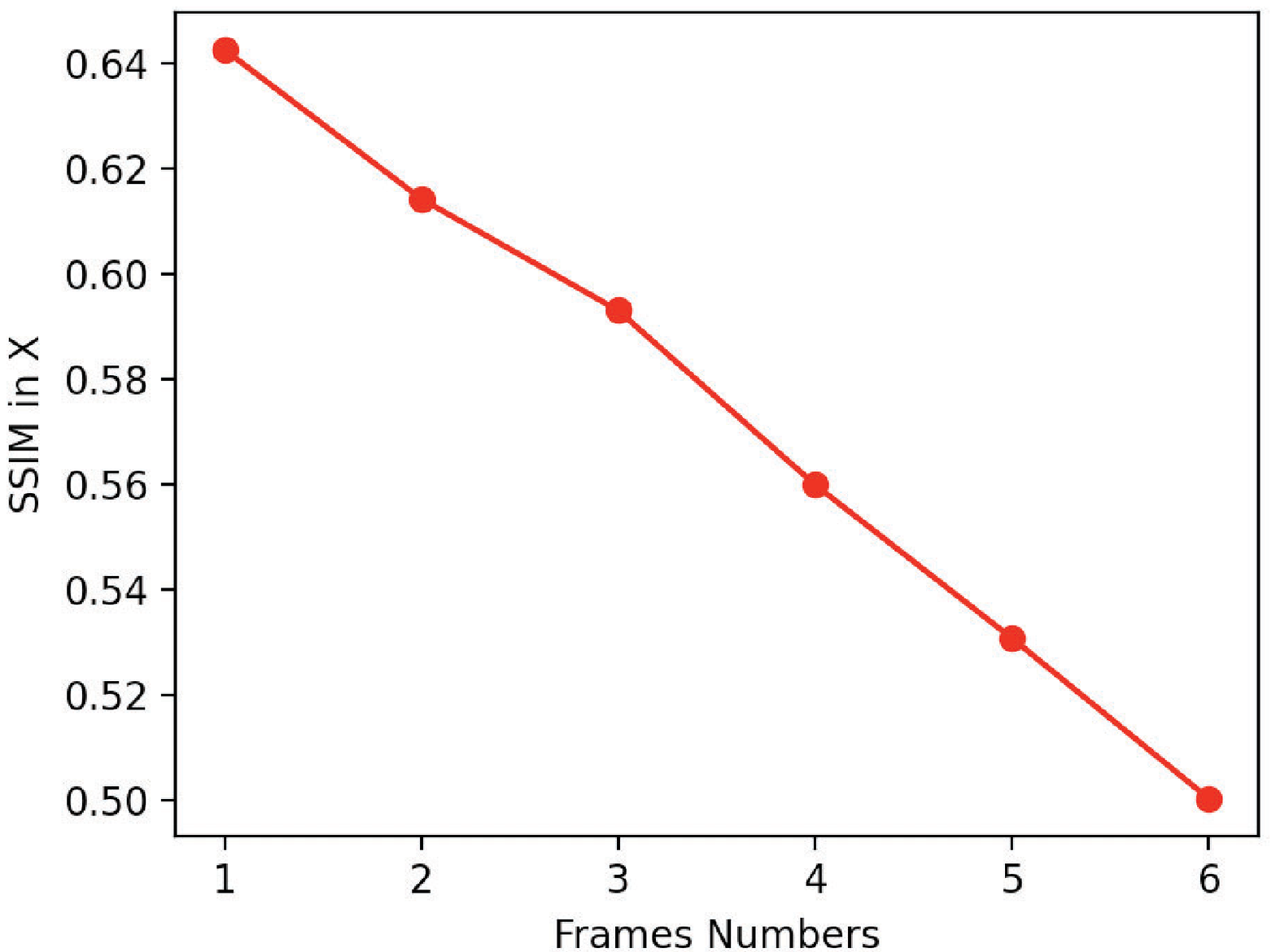}
      \caption{Average of SSIM in Longitude(X)}
   \end{subfigure}
   \begin{subfigure}{0.35\textwidth}
      \centering
      \includegraphics[width=\textwidth]{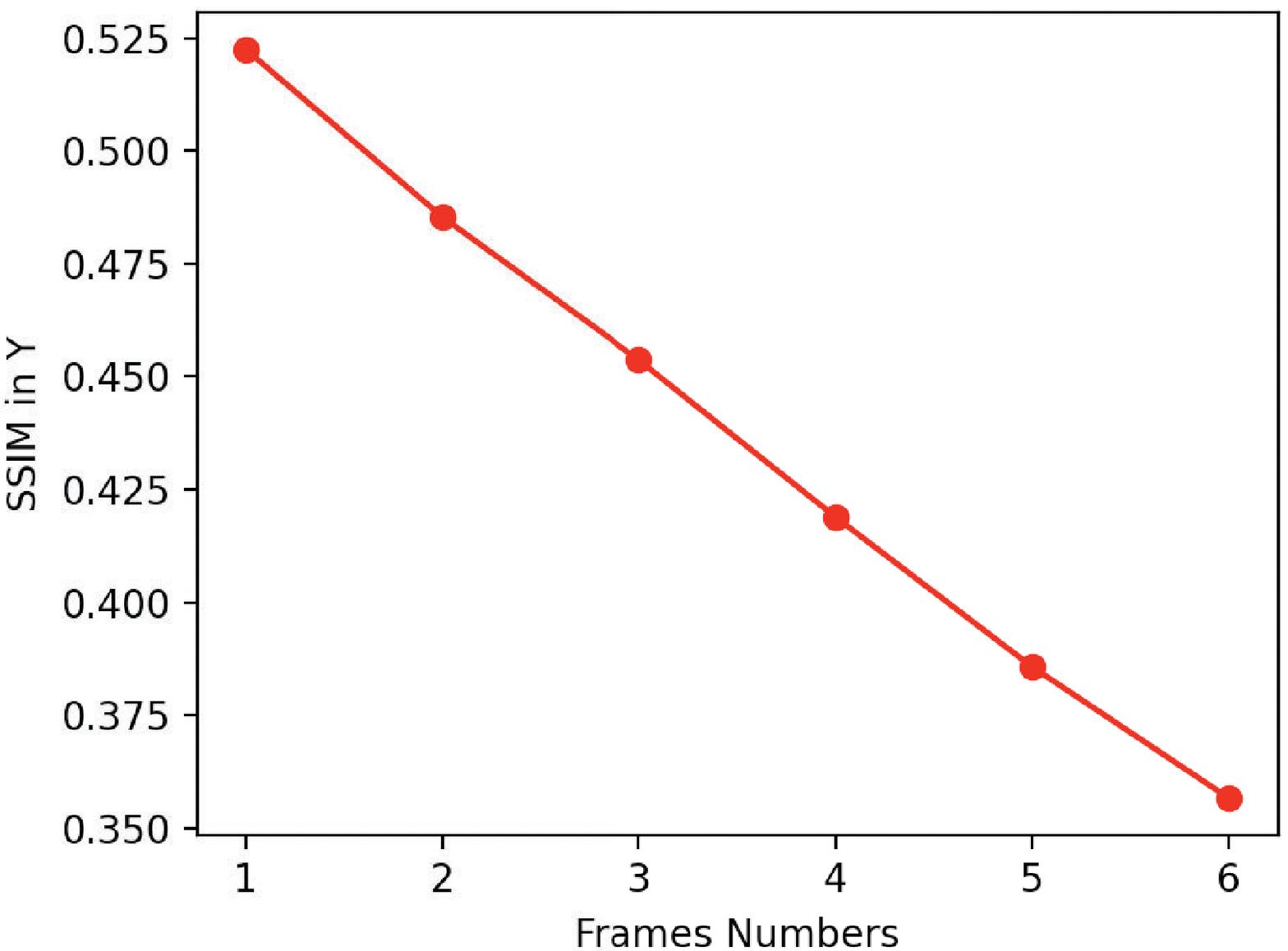}
      \caption{Average of SSIM in Latitude(Y)}
   \end{subfigure}

   \begin{subfigure}{0.35\textwidth}
      \centering
      \includegraphics[width=\textwidth]{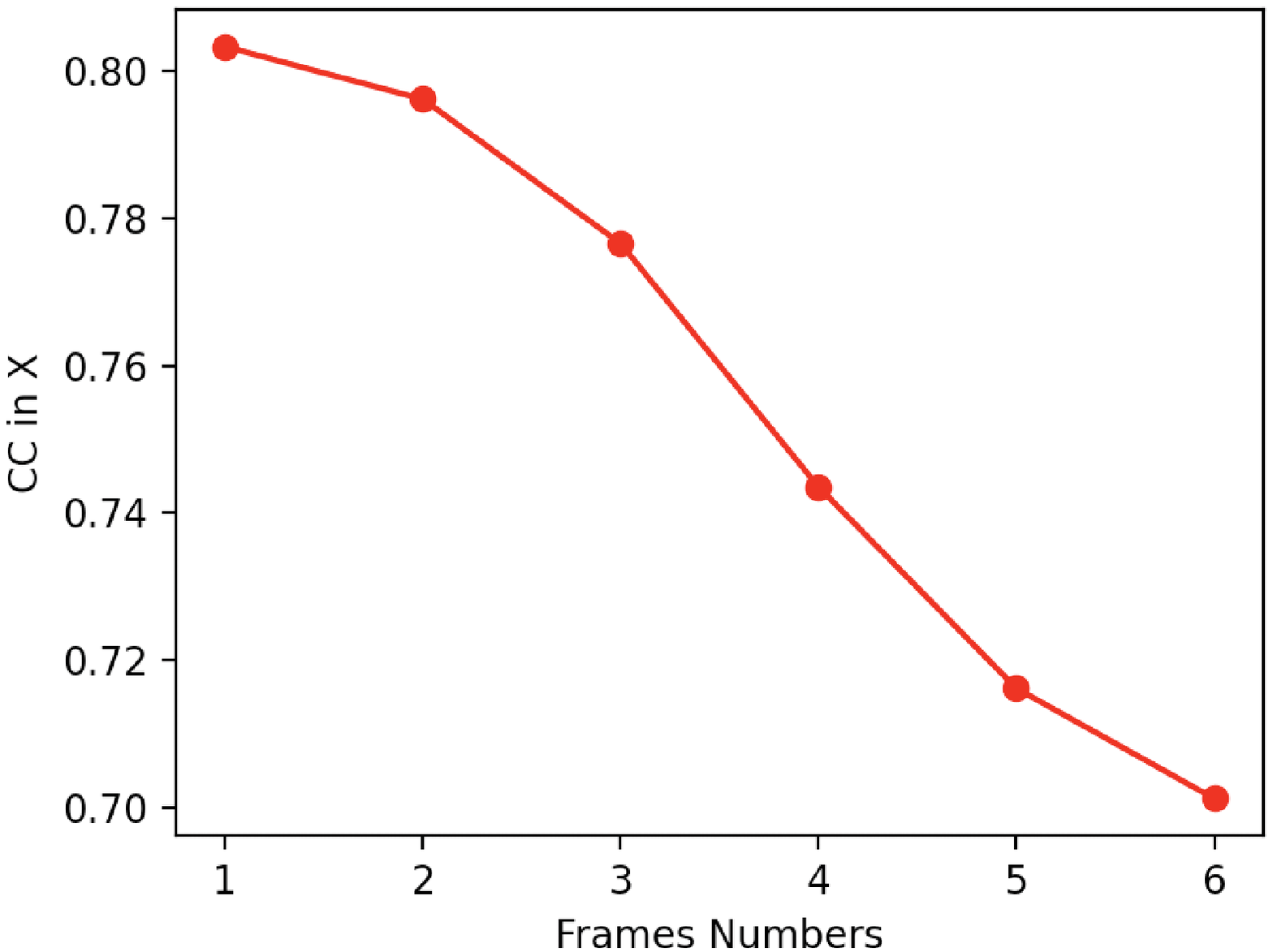}
      \caption{Average of CC in Longitude(X)}
   \end{subfigure}
   \begin{subfigure}{0.35\textwidth}
      \centering
      \includegraphics[width=\textwidth]{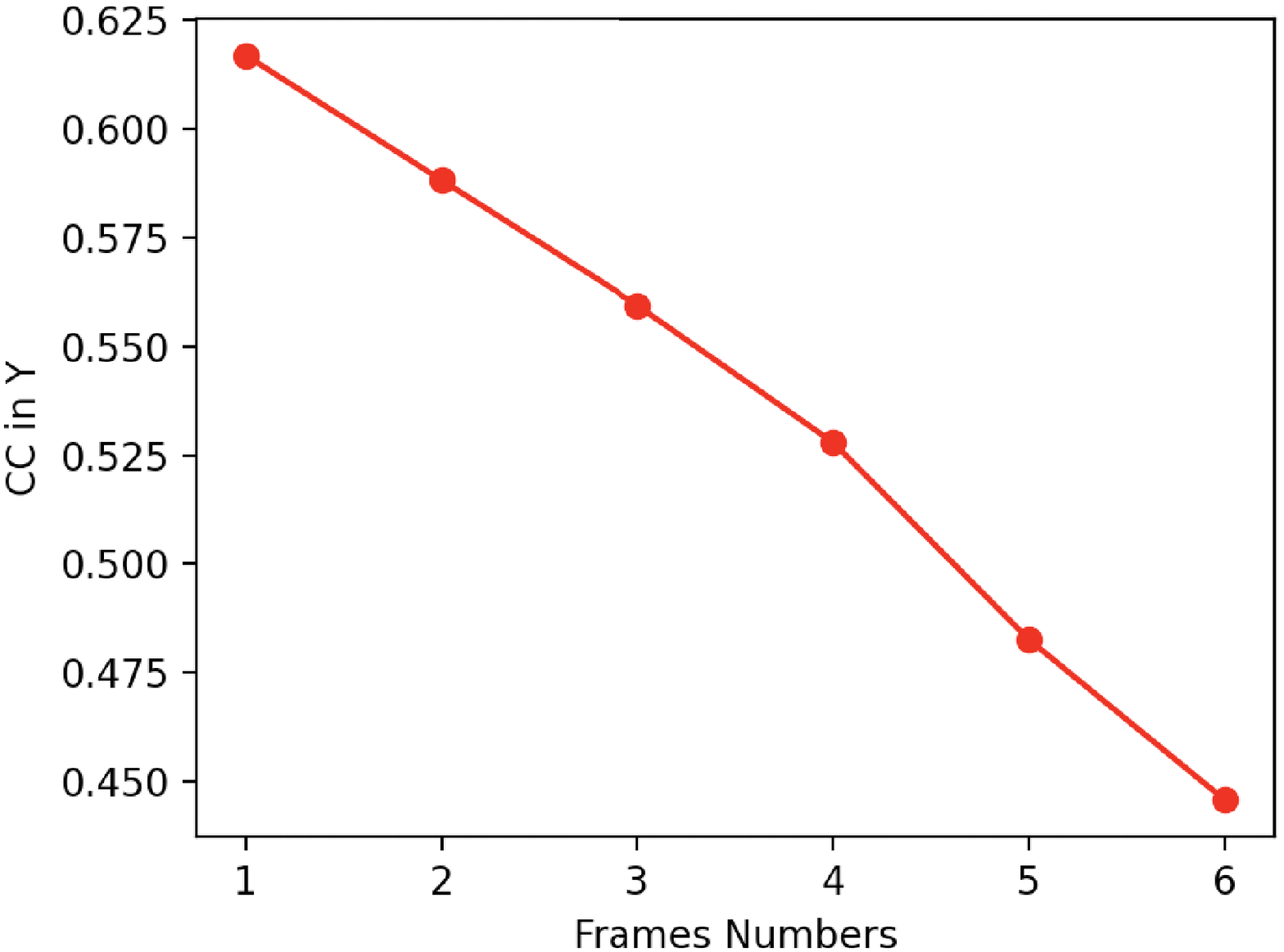}
      \caption{Average of CC in Latitude(Y)}
   \end{subfigure}
   \caption{Average CC and SSIM of the Speed Diagrams on the Whole Testing Set}
\end{figure}

After carefully observing the magnetic field sequences of the training set and testing set, we find that the evolution in the longitude direction is obviously faster than that in the latitude direction in most sequences. The speed magnitude of the magnetic field structure in the longitude direction is significantly larger than that in the latitude direction. There are more cases where the positive and negative magnetic field structures are separated in the longitude direction. In general, the movement speed in the longitude direction is more obvious, which can reflect the large-scale change of the magnetic field structure movement in the evolution. Based on these comparisons of the speed diagrams and the average SSIM and CC of speed diagrams on the whole testing set above, we believe that the large-scale movement speed of the predicted results is generally consistent with the real magnetic field images.

\section{Conclusions and Discussions}
\label{sect:discussion}
This paper makes efforts to predict the short-term, large-scale evlution of the photospheric magnetic fields. On the basis of the constructed magnetic field data in active regions, a spatiotemporal LSTM neural network are used to build our model. Experimental results show that the model can roughly predict the large-scale evolution of the photospheric magnetic field in active regions in the next 6 hours. Specifically, we can draw the following conclusions: (1) Through the statistical learning of a large number of magnetic field sequences covering various evolution situations, The model learned a prediction pattern which can be applied to predict the evolution of new magnetic field sequences in next 6 hour. The predicted results are generally consistent with the real magnetic field in terms of large-scale magnetic field structure and movement speed. On the other hand, this also verifies that the evolution of a large number of magnetic field sequences in time and space has certain certainty and regularity, which could be roughly characterized by the used neural network model. (2) The performance of the model is affected by the prediction time, the shorter the prediction time, the higher the accuracy of the predicted magnetic field evolution. (3) The performance of the model is stable not only for active regions in the north and south but also for data in positive and negative regions.

During the experiment, we also found some aspects of the prediction results that might be controversial and some limitations of the model. Our following analyses and discussions about these parts will provide valuable ideas and references for subsequent in-depth research.

\begin{figure}[htbp]
   \label{fig:example}
   \begin{subfigure}{\textwidth}
      \centering
      \includegraphics[width=\textwidth]{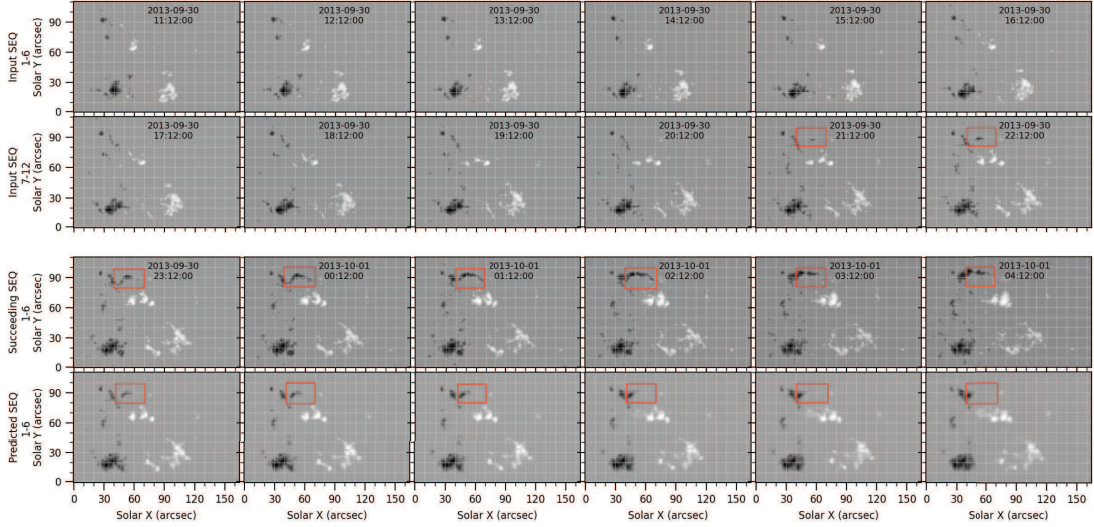}
      \caption{One Group of Predicted Results when Magnetic Flux Emergence is Not Predicted}
   \end{subfigure}

   \begin{subfigure}{\textwidth}
      \centering
      \includegraphics[width=\textwidth]{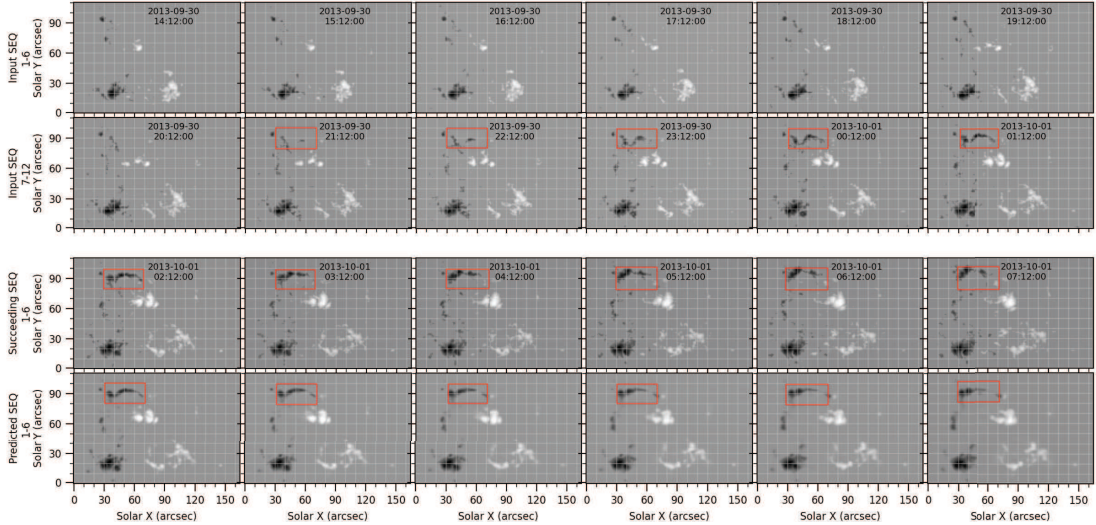}
      \caption{One Group of Predicted Results When Magnetic Flux Emergence is Predicted}
   \end{subfigure}
   \caption{Predicted Results about Magnetic Flux Emergence}
\end{figure}
The magnetic flux emergence is one of the important phenomena in the evolution of AR magnetic field. We select two groups of predicted results from AR 11855 for display: one group predicted the emerging magnetic field block, and the other group failed to predict the emerging magnetic field block. As shown in the red box of Figure 7a, the new emerging magnetic field block appears in the 11th frame of the input sequence, and there is no emerging magnetic field block in the predicted results. In Figure 7b, a small emerging block of magnetic flux appears in the 8th frame of the input sequence, and this block continues to perform in frames 9, 10, 11 and 12 of the input sequence. In the predicted results, this emerging magnetic field block finally appeares. We see the following two phenomena by comparing more similar examples about whether the emergence of new magnetic flux patches can be predicted. When the new magnetic field structure emerges in the 11th to 12th frames of the input sequence, the predicted results are very likely to be unable to predict the emerging block. When a new magnetic field structure emerges in the 8th to 10th frame and earlier in the input sequence, there is a high probability that the emerging block will appear in the predicted results. Accordingly, we can conclude that when the new magnetic field structure emerges earlier and performs a certain performance in the input sequence, the model could predict this emerging block with a high probability.

\begin{figure}[htbp]
   \label{fig:example}
   \begin{subfigure}{\textwidth}
      \centering
      \includegraphics[width=\textwidth]{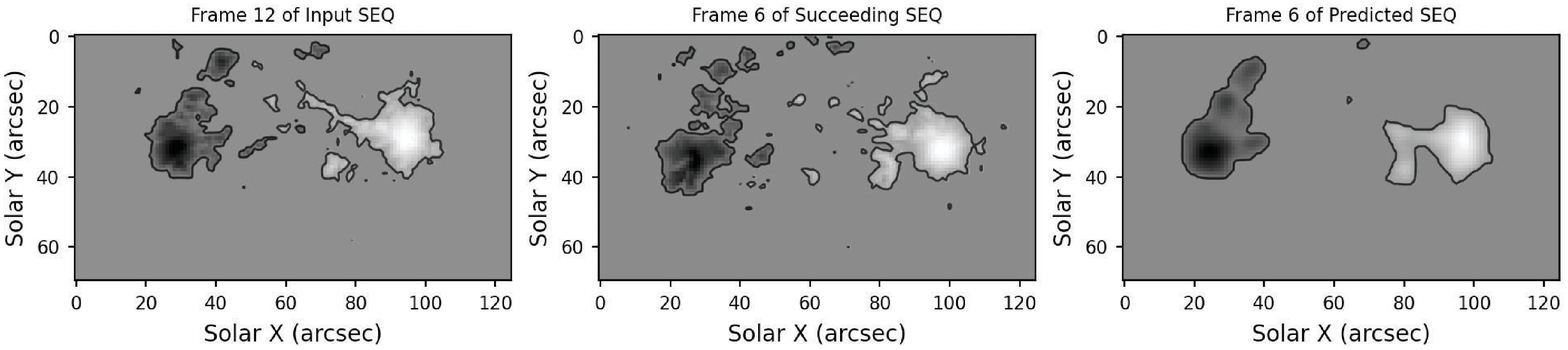}
      \caption{One Group of Contour Diagrams on AR 11824}
   \end{subfigure}

   \begin{subfigure}{\textwidth}
      \centering
      \includegraphics[width=\textwidth]{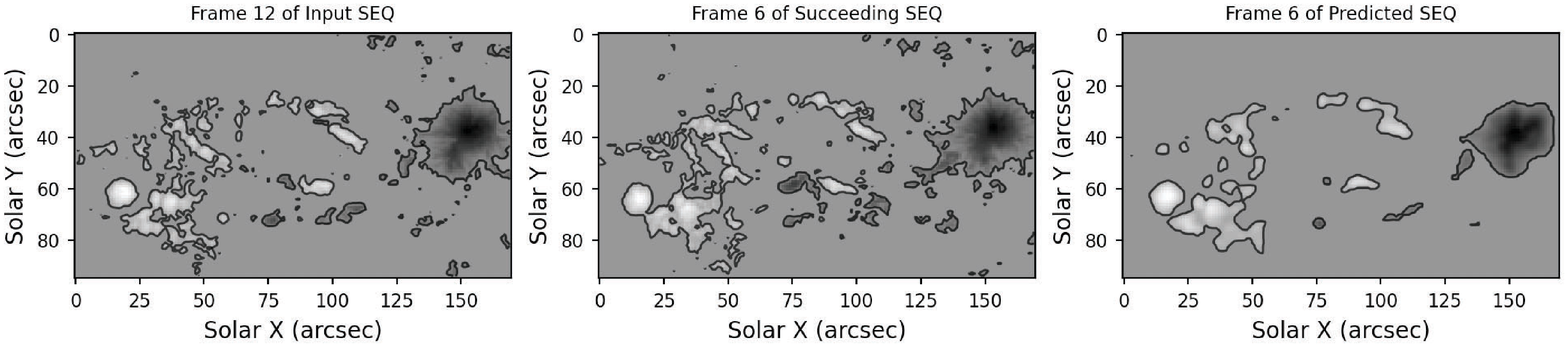}
      \caption{One Group of Contour Diagrams on AR 12219}
   \end{subfigure}
   \caption{The contour diagrams about tow groups of predicted results are shown. In panels(a-b), the first image is the 12th frame of the input sequence, the second image is the real magnetic field image of the 6th hour (frame) after input sequence, and the third image is the predicted magnetic field image of the 6th hour.}
\end{figure}

The model we trained is mainly able to predict the large-scale evolution of the photospheric magnetic field in solar active regions. When conducting researches on the fine structure of the magnetic field based on the predicted results, there is a big deviation. Two groups of predicted results were selected from AR 11824 and AR 12219, and their contour diagrams are shown in Figure 8. We can see that, compared with the 12th magnetic field images of the input sequence, the position of the real magnetic field of the 6th frame in the succeeding sequence has changed, and the structure has also been deformed. In the sixth frame of the predicted results, the large-scale displacement and deformation are roughly consistent with the real magnetic field image, but there are still considerable deviations in the deformation of fine structure. The main reason is that large-scale structure deformation and movement are relatively easier to be characterized by the used network. The relationships in deformations of fine structure are more complex, and their patterns are more difficult to be learned. We have carefully checked the predicted results of algorithms in the field of computer vision (\citealt{kwon_predicting_2019_3} , \citealt{wang_memory_2019_4}), and find that there is also the problem that the deformations of the predicted results on fine structure is less consistent with the real video. 

\begin{figure}[htbp]
   \label{fig:example}
   \begin{subfigure}{\textwidth}
      \centering
      \includegraphics[width=\textwidth]{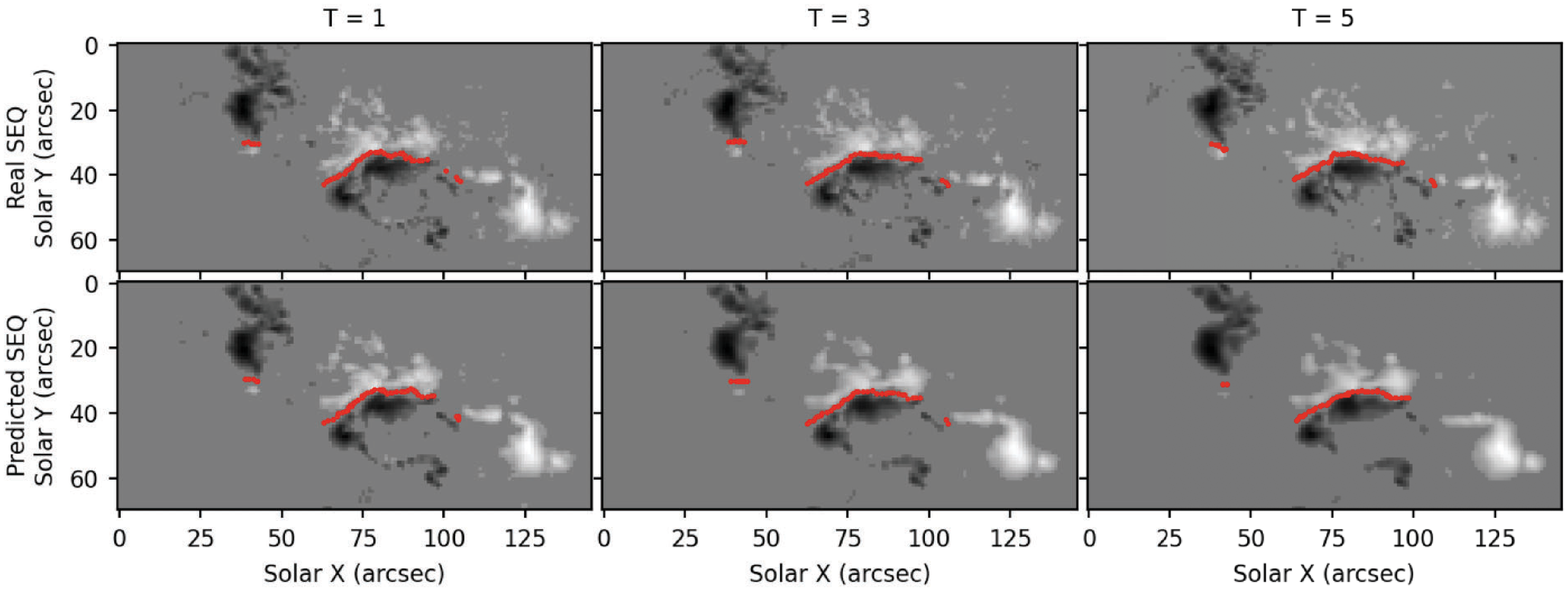}
      \caption{One Group of Predicted Results On AR 11158}
   \end{subfigure}
   \begin{subfigure}{\textwidth}
      \centering
      \includegraphics[width=\textwidth]{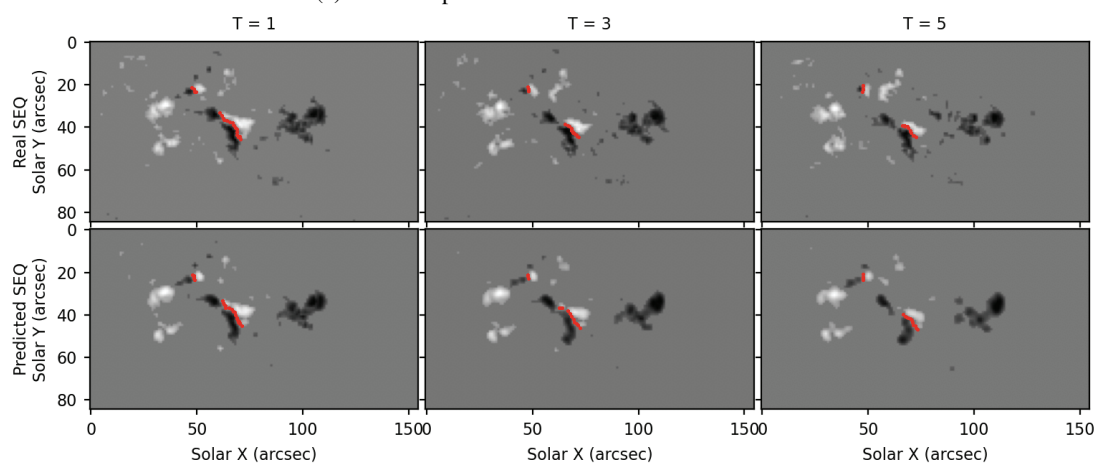}
      \caption{One Group of Predicted Results On AR 12089}
   \end{subfigure}
   \caption{Two groups of predicted results with the magnetic neutral lines are shown. The red line is the magnetic neutral line drawn by us. The first row is the real magnetic field images, and the second row is the corresponding predicted results. T = 1, T = 2, T = 3 indicate that the time is 1 hour, 3 hours and 5 hours after input sequence respectively.}
\end{figure}

The magnetic neutral line is the line that separates the opposite magnetic polarity regions (positive and negative) in photosphere magnetic field and is often located in the place with a large gradient of the magnetic field images. We pick out two groups of predicted results in which the positive and negative regions are relatively close, and their approximate magnetic neutral lines are drawn in Figure 9. It can be seen that the predicted magnetic neutral lines are roughly consistent with the neutral lines of real magnetic images in the position, length, and inclination angle. We also compare more groups of predicted results during the experiment and find that when the positive and negative regions of magnetic fields are close, the drawn magnetic neutral lines on predicted results have good consistency with the real magnetic neutral lines. When the positive and negative regions are far away, the drawn magnetic neutral lines on predicted results have deviations with the real magnetic neutral lines. Because when the positive and negative regions of magnetic fields are close to each other, The magnetic neutral lines can better indicate some solar activities such as flares and filaments. So we could conclude that the predicted results are roughly consistent with the real magnetic field images in the large-scale structure and evolution of magnetic neutral lines.

Like other deep learning models, our model is also data-driven, and its performance is affected by whether the training data is sufficient. Although the training data we used has tried to cover as much magnetic field evolution as possible, the amount of data is still not large enough compared with the complicated and changeable evolution. The physical evolution contained in the data is still not comprehensive enough. Also, the limitation of the computing power of the equipment and the slow operation efficiency of the network also affects the model's performance. Therefore, under the current data level, the predicted results by our model could not be applied to the study of magnetic field on fine structures.

In conclusion, we have constructed a prediction model based on a spatiotemporal LSTM neural network to realize the short-term prediction of the large-scale evolution of the photospheric magnetic field in solar active regions. In the future, we will try to solve these above controversial aspects and limitations, and research new magnetic field prediction models with higher prediction accuracy and longer prediction time.

\begin{acknowledgements}
This study is supported by the National Natural Science Foundation of China (12073077,11873027,U2031140,11773072,12063002). We thank the SDO science teams for providing the excellent data.
\end{acknowledgements}

\label{lastpage}
\bibliographystyle{raa}
\bibliography{Re}

\begin{thebibliography}{21}
\providecommand\natexlab[1]{#1}
\providecommand\JournalTitle[1]{#1}

\bibitem[Ba {et~al.}(2016)]{ba_layer_2016_15}
Ba, J.~L., Kiros, J.~R., \& Hinton, G.~E. 2016, arXiv: 1607.06450

\bibitem[Bengio {et~al.}(2015)]{bengio_scheduled_nodate_19}
Bengio, S., Vinyals, O., Jaitly, N., \& Shazeer, N. 2015, in Advances in Neural
  Information Processing Systems, ed. C.~Cortes, N.~Lawrence, D.~Lee,
  M.~Sugiyama, \& R.~Garnett, Vol.~28 (Curran Associates, Inc.), 1171

\bibitem[Bobra {et~al.}(2014)]{bobra_helioseismic_2014_13}
Bobra, M.~G., Sun, X., Hoeksema, J.~T., {et~al.} 2014, Solar Physics, 289, 3549

\bibitem[Covas(2020)]{covas_transfer_2020_9}
Covas, E. 2020, Astronomische Nachrichten, 341, 384

\bibitem[Covas {et~al.}(2019)]{covas_neural_2019_8}
Covas, E., Peixinho, N., \& Fernandes, J. 2019, Solar Physics, 294, 24

\bibitem[{Dani} \& {Sulistiani}(2019)]{dani_prediction_2019_7}
{Dani}, T., \& {Sulistiani}, S. 2019, in Journal of Physics Conference Series,
  Vol. 1231, Journal of Physics Conference Series, 012022

\bibitem[Farnebäck(2003)]{goos_two-frame_2003_20}
Farnebäck, G. 2003, in the 13th Scandinavian conference on Image analysis, ed.
  J.~Bigun \& T.~Gustavsson, Vol. 2749 (Berlin, Heidelberg: Springer Berlin
  Heidelberg), 363

\bibitem[Getachew(2019)]{getachew_spatial-temporal_nodate_11}
Getachew, T. 2019, Spatial-temporal structure and distribution of the solar
  photospheric magnetic field (Oulu : Oulun yliopisto)

\bibitem[Huang {et~al.}(2018)]{Huang_2018}
Huang, X., Wang, H., Xu, L., {et~al.} 2018, The Astrophysical Journal, 856, 7

\bibitem[Kingma \& Ba(2017)]{kingma_adam_2017_18}
Kingma, D.~P., \& Ba, J. 2017, in International Conference on Learning
  Representations (San Diego, CA: Springer)

\bibitem[Kwon \& Park(2019)]{kwon_predicting_2019_3}
Kwon, Y.-H., \& Park, M.-G. 2019, in 2019 {IEEE}/{CVF} {Conference} on
  {Computer} {Vision} and {Pattern} {Recognition} ({CVPR}) (Long Beach, CA,
  USA: IEEE), 1811

\bibitem[Li {et~al.}(2009)]{li_image_2009_14}
Li, H., Zhang, H., Guo, X., \& Hu, G. 2009, Tsinghua Science and Technology,
  14, 541

\bibitem[Nishizuka {et~al.}(2018)]{nishizuka_deep_2018_6}
Nishizuka, N., Sugiura, K., Kubo, Y., Den, M., \& Ishii, M. 2018, \apj, 858,
  113

\bibitem[Oprea {et~al.}(2020)]{oprea_review_2020_1}
Oprea, S., Martinez-Gonzalez, P., Garcia-Garcia, A., {et~al.} 2020, arXiv:
  2004.05214

\bibitem[Pala \& Atici(2019)]{pala_forecasting_2019_5}
Pala, Z., \& Atici, R. 2019, Solar Physics, 294, 50

\bibitem[Paszke {et~al.}(2019)]{paszke_pytorch_nodate_17}
Paszke, A., Gross, S., Massa, F., {et~al.} 2019, in Advances in Neural
  Information Processing Systems, ed. H.~Wallach, H.~Larochelle,
  A.~Beygelzimer, F.~d\textquotesingle Alch\'{e}-Buc, E.~Fox, \& R.~Garnett,
  Vol.~32 (Curran Associates, Inc.), 8026

\bibitem[Pesnell {et~al.}(2012)]{pesnell_solar_2012_12}
Pesnell, W.~D., Thompson, B.~J., \& Chamberlin, P.~C. 2012, Solar Physics, 275,
  3

\bibitem[Shi {et~al.}(2015)]{shi_convolutional_nodate_2}
Shi, X., Chen, Z., Wang, H., {et~al.} 2015, in Advances in Neural Information
  Processing Systems, ed. C.~Cortes, N.~Lawrence, D.~Lee, M.~Sugiyama, \&
  R.~Garnett, Vol.~28 (Curran Associates, Inc.), 802

\bibitem[{Wang} {et~al.}(2019)]{wang_memory_2019_4}
{Wang}, Y., {Zhang}, J., {Zhu}, H., {et~al.} 2019, in 2019 IEEE/CVF Conference
  on Computer Vision and Pattern Recognition (CVPR), 9146

\bibitem[Wiegelmann {et~al.}(2014)]{wiegelmann_magnetic_2014_10}
Wiegelmann, T., Thalmann, J.~K., \& Solanki, S.~K. 2014, Astronomy and
  Astrophysics Review, 22, 78

\bibitem[Wu \& He(2018)]{wu_group_2018_16}
Wu, Y., \& He, K. 2018, in Proceedings of the European Conference on Computer
  Vision (ECCV), ed. V.~Ferrari, M.~Hebert, C.~Sminchisescu, \& Y.~Weiss,
  Vol.~16 (Springer)

\end{thebibliography}
\end{document}